\documentclass[twocolumn,trackchanges]{aastex62} 

\shorttitle{Constraining the dark matter-vacuum energy interaction using the EDGES 21-cm absorption signal}
\shortauthors{Wang et al.}

\def\ie{{\frenchspacing\it i.e.}}
\def\eg{{\frenchspacing\it e.g.~}}

\def\be{\begin{equation}}
\def\ee{\end{equation}}
\def\ba{\begin{eqnarray}}
\def\ea{\end{eqnarray}}

\begin{document}

\title{Constraining the dark matter-vacuum energy interaction using the EDGES 21-cm absorption signal}

\correspondingauthor{Yuting Wang, Gong-Bo Zhao}
\email{ytwang@nao.cas.cn, gbzhao@nao.cas.cn}

\author[0000-0001-7756-8479]{Yuting Wang}
\affiliation{National Astronomy Observatories, Chinese Academy of Science, Beijing, 100101, P.R.China}

\author[0000-0003-4726-6714]{Gong-Bo Zhao}
\affiliation{National Astronomy Observatories, Chinese Academy of Science, Beijing, 100101, P.R.China}
\affiliation{University of Chinese Academy of Sciences, Beijing, 100049, P.R.China}
\affiliation{Institute of Cosmology and Gravitation, University of Portsmouth, Portsmouth, PO1 3FX, UK}



\begin{abstract}
The recent measurement of the global 21-cm absorption signal reported by the Experiment to Detect the Global Epoch of Reionization Signature (EDGES) Collaboration is in tension with the prediction of the $\Lambda$CDM model at a $3.8\,\sigma$ significance level. In this work, we report that this tension can be released by introducing an interaction between dark matter and vacuum energy. We perform a model parameter estimation using a combined dataset including EDGES and other recent cosmological observations, and find that the EDGES measurement can marginally improve the constraint on parameters that quantify the interacting vacuum, and that the combined dataset favours the $\Lambda$CDM at 68\% CL. This proof-of-the-concept study demonstrates the potential power of future 21-cm experiments to constrain the interacting dark energy models.
\end{abstract}

\keywords{Cosmology: dark energy}

\section{Introduction} \label{sec:intro}
Recently, the Experiment to Detect the Global Epoch of Reionization Signature (EDGES) Collaboration reported an excess 21-cm absorption signal at the effective redshift $z\sim17$ \citep{Bowman:2018}. The amplitude of this observed signal is $T_{\rm 21}=-500^{+200}_{-500}$ mK, where the error, including potential systematic uncertainties, is at the $99\%$ confidence level (CL) \citep{Bowman:2018}. Surprisingly, this measurement is in tension with the theoretical prediction in the standard $\Lambda$CDM cosmology at about a $3.8\,\sigma$ significance level, namely, the measured $T_{\rm 21}$ almost doubles the $\Lambda$CDM prediction, which is  $T_{\rm 21}=-209$ mK \citep{Barkana:2018lgd}. 

Much attention from the astrophysics community has been attracted to this discovery, and various interpretations have been proposed to explain the discrepancy. As $T_{\rm 21} \propto [1- T_{\rm CMB}(z)/T_{\rm S}(z)]/H(z)$, where $T_{\rm 21}$ is the measured intensity of the 21-cm radiation relative to the Cosmic Microwave Background (CMB) temperature $T_{\rm CMB}(z)$, $T_{\rm S}(z)$ is the spin temperature of the hydrogen gas, and $H(z)$ the Hubble parameter, there are in principle three ways (and their combinations) to make $T_{\rm 21}$ more negative to be compatible with the EDGES measurement: (A) reduce the spin temperature $T_{\rm S}$ by introducing new cooling mechanisms, \eg, the dark matter-baryon scattering \citep{Barkana:2018lgd, Fialkov:2018xre, Munoz:2018pzp, Berlin:2018sjs}; or (B) raise $T_{\rm CMB}$ by additional radio background \citep{Feng:2018rje, Ewall-Wice:2018bzf, Fraser:2018acy}; or (C) reduce the Hubble parameter \citep{Costa:2018aoy,Hill:2018lfx}.

In this paper, we propose to release the tension by reducing the Hubble parameter through the interaction between dark matter and dark energy. Specifically, we consider the interacting vacuum energy model\footnote{The idea of decaying vacuum energy is a recurring concept to explain the accelerating expansion of the Universe \citep{Bertolami:1986bg,Freese:1986dd,Chen:1990jw,Carvalho:1991ut,Berman:1991zz,Pavon:1991uc,AlRawaf:1995rs,Shapiro:2000dz,Sola:2011qr}.} proposed in \citep{Wands:2012vg}, and perform a parameter estimation for this model using a joint dataset including EDGES and other kinds of recent cosmological measurements. 

This paper is organised as follows. In the next section, we present a brief description of the 21-cm absorption observable. Then we introduce the interacting vacuum energy model in Sec. \ref{sec:model}, before showing the cosmological constraint on the interacting vacuum energy model using observations in Sec. \ref{sec:test}. The last section is devoted to conclusion and discussions.

\section{The 21-cm absorption signal}
At the Rayleigh-Jeans limit, the brightness temperature of the observed radiation field is \citep{Field:1958, Furlanetto:2006jb},
\ba
T_b (z, \nu) = T_{\rm CMB} (z) e^{-\tau_\nu} +T_{\rm S}(z)\left(1-e^{-\tau_\nu}\right)\,, 
\ea where $z$ and $\nu$ denote redshift and frequency respectively, and $\tau_\nu$ is the optical depth of the inter-galactic medium at frequency $\nu$. $T_S(z)$ is the spin temperature, and $T_{\rm CMB}$ the CMB temperature which evolves with redshift as $T_{\rm CMB} (z) = 2.725 (1+z)$ K.

The 21-cm signal is caused by the hyperfine splitting of neutral hydrogen atoms. The transition from the triplet state to the singlet state corresponds to the emission of photons of wavelength at 21 cm, whose frequency is $\nu_0=1420.4$ MHz. The intensity of the 21-cm radiation relative to the CMB temperature is thus,
\ba
T_{\rm 21}(z) &\approx& \frac{T_{\rm S}(z) - T_{\rm CMB}(z)}{1+z}\tau_{\nu_0}(z)\,, \label{eq:T21}\\
\tau_{\nu_0}(z) &=&  \frac{2c^3\hbar A_{10}n_{\rm HI}}{16k_{\rm B}\nu_0^2T_{\rm S}(z)H(z)}\,, \label{eq:tau}
\ea
where $c$ is the speed of light, $\hbar$ the reduced Planck constant, and $k_{\rm B}$ the Boltzmann constant. $A_{10}=2.85\times10^{-15} s^{-1}$ is the emission coefficient of the  spontaneous transition from the triplet state to the singlet state. $n_{\rm HI}$ is the number density of neutral hydrogen, and $H(z)$ the Hubble parameter as a function of redshift. In $\Lambda$CDM, $H(z)$ can be approximated as $H_0\sqrt{\Omega_{\rm M}(1+z)^3}$ at $z\gg1$. Such that the optical depth in Eq. (\ref{eq:tau}) can be rewritten as \citep{Furlanetto:2006jb},
\ba
\tau_{\nu_0}^{\rm \Lambda CDM} &\approx& 8.6\times10^{-3}x_{\rm HI} \left[ \frac{T_{\rm CMB}(z)}{T_{\rm S}(z)}\right] \left(\frac{\Omega_bh^2}{0.02}\right) \nonumber \\ 
&\times& \left[ \left( \frac{0.15}{\Omega_{\rm M}h^2}\right)\left( \frac{1+z}{10}\right)\right]^{1/2} \ .
\ea
where $x_{\rm HI}$ is the neutral hydrogen fraction, $\Omega_bh^2$ and $\Omega_{\rm M}h^2$ are the physical baryon and matter density respectively.

\section{The interacting vacuum energy model} \label{sec:model}

\begin{figure}[t!]   
\includegraphics[scale=0.3]{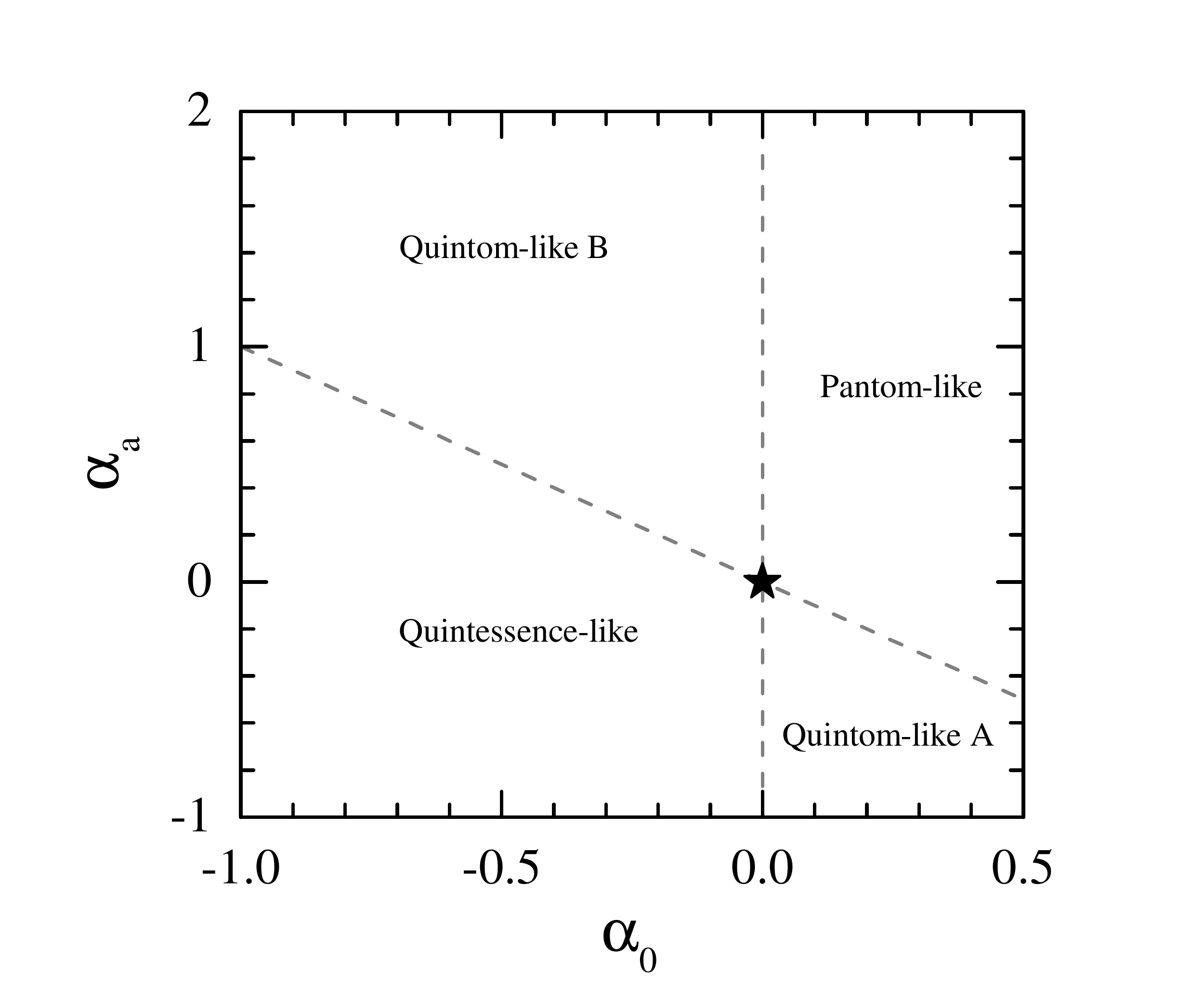}
\caption{The dashed lines of $\alpha_0=0$ and $\alpha_0+\alpha_a=0$ divide the parameter space of $\alpha_0$ and $\alpha_a$ into four regions, where $w_V^{\rm eff}$ is greater than $-1$ in the past and smaller than $-1$ today in the models of  ``Quintom-like A", while $w_V^{\rm eff}$ crosses $-1$ from the values smaller than $-1$ to that greater than $-1$ in the models of  ``Quintom-like B". The black star denotes the $\Lambda$CDM model.}\label{fig:scalar_field_like}
\end{figure} 

With presence of the interacting vacuum energy, the continuity equations for the interacting vacuum $V$ and dark matter ${\rho}_{\rm dm}$ are,
\ba
\dot{V} &=& Q\,, \nonumber\\
\label{eq:Vrho}\dot{\rho}_{\rm dm} +3 H \rho_{\rm dm} &=& -Q\,,
\ea
where $Q$ is the interaction between the vacuum energy and dark matter. Vacuum energy has a non-varying equation of state, \ie\, $w\equiv-1$, but a time-evolving energy density due to the interacting term $Q$, which is different from the interacting dark energy models discussed in \citep{Costa:2018aoy}, where the equation of state of dark energy is a constant $w$, but $w\neq-1$. In this work, we consider an interaction of the form, 
\ba\label{eq:Q}
Q &=& 3 \alpha H \frac{ \rho_{\rm dm} V}{\rho_{\rm dm}+V}\,, \ea
Note that $\alpha$ can be a function of time in general. In this work, we parameterise the time-dependence as, \ba\label{eq:alpha}\alpha(a) &=& \alpha_0+\alpha_a(1-a) \,.\ea As shown, $\alpha$ approaches $\alpha_0$ and $\alpha_0+\alpha_a$ in limits of $a=1$ and $a=0$ respectively, and interpolates linearly in between. In this model, $\alpha_a=-{\rm d}\alpha/{\rm d} a$, thus it can be used as an indicator of the dynamics of vacuum. The effective equation of state for vacuum energy is
\ba
w_V^{\rm eff} = -1 - \alpha(a)\frac{\rho_{\rm dm}}{\rho_{\rm dm}+V}\,.
\ea
This parametrization of $\alpha(a)$ can realize a quintom-like effective dark energy with the EoS crossing $-1$, as shown in the upper left and the lower right regions of Fig. \ref{fig:scalar_field_like}. In contrast, a constant interaction parameter, \ie\, $\alpha(a)=\alpha_0$ discussed in \citep{Wang:2013qy, Wang:2014xca}, only can yield a quintessence-like dark energy with $w_V^{\rm eff}>-1$ for a negative $\alpha_0$ or a phantom-like dark energy with $w_V^{\rm eff}<-1$ for a positive $\alpha_0$.

The Friedmann equation reads,
\ba\label{eq:H}
H^2=\frac{8 \pi G}{3} \left[\rho_b+\rho_r+\rho_{\rm dm}+V\right] \,, \label{eq:Hz} 
\ea
where baryons and radiation follow the standard conservation equations. The expansion history of the Universe can be solved by combining Eqs (\ref{eq:Vrho}), (\ref{eq:Q}), (\ref{eq:alpha}) and (\ref{eq:H}). Apparently, any nonzero $\alpha$ yields a modification of expansion history compared with that in the $\Lambda$CDM model \citep{Wang:2013qy}. As the 21-cm temperature $T_{\rm 21}$ depends on $\tau_{\nu0}$, which further depends on the expansion rate $H$ through Eq (\ref{eq:tau}), the interacting vacuum model can leave imprints on the 21-cm observables.

At the perturbation level, we consider an energy flow that is parallel to the 4-velocity of dark matter, \ie\, $Q^\mu_{\rm dm}=-Qu^\mu_{\rm dm}$. In this case dark matter particles follow geodesics as in $\Lambda$CDM, but the continuity equation gets modified, namely, the velocity perturbation for dark matter is not affected by the interaction and obeys the standard equation \citep{Wang:2013qy, Wang:2014xca}
\ba
\dot{\theta}_{\rm dm}=0 \,.
\ea
Thus we will evolve the perturbation equations in a synchronous gauge that is comoving with the dark matter\footnote{If
the initial value of $\theta_{\rm dm}$ is set to zero, it would remain zero at all times.}. Meanwhile, the dark matter density contrast ${\delta}_{\rm dm}$ evolves in the dark matter-comoving frame \citep{Wang:2013qy, Wang:2014xca},
\ba
\dot{\delta}_{\rm dm} = - \frac{\dot h}{2} + \frac{Q}{\rho_{\rm dm}} \delta_{\rm dm} \,,
\ea
where $h$ is the scalar mode of metric perturbations in the synchronous gauge. In this gauge, the vacuum energy is spatially homogeneous, \ie\,$\delta V=0$.

\begin{figure}[b!]   
\includegraphics[scale=0.3]{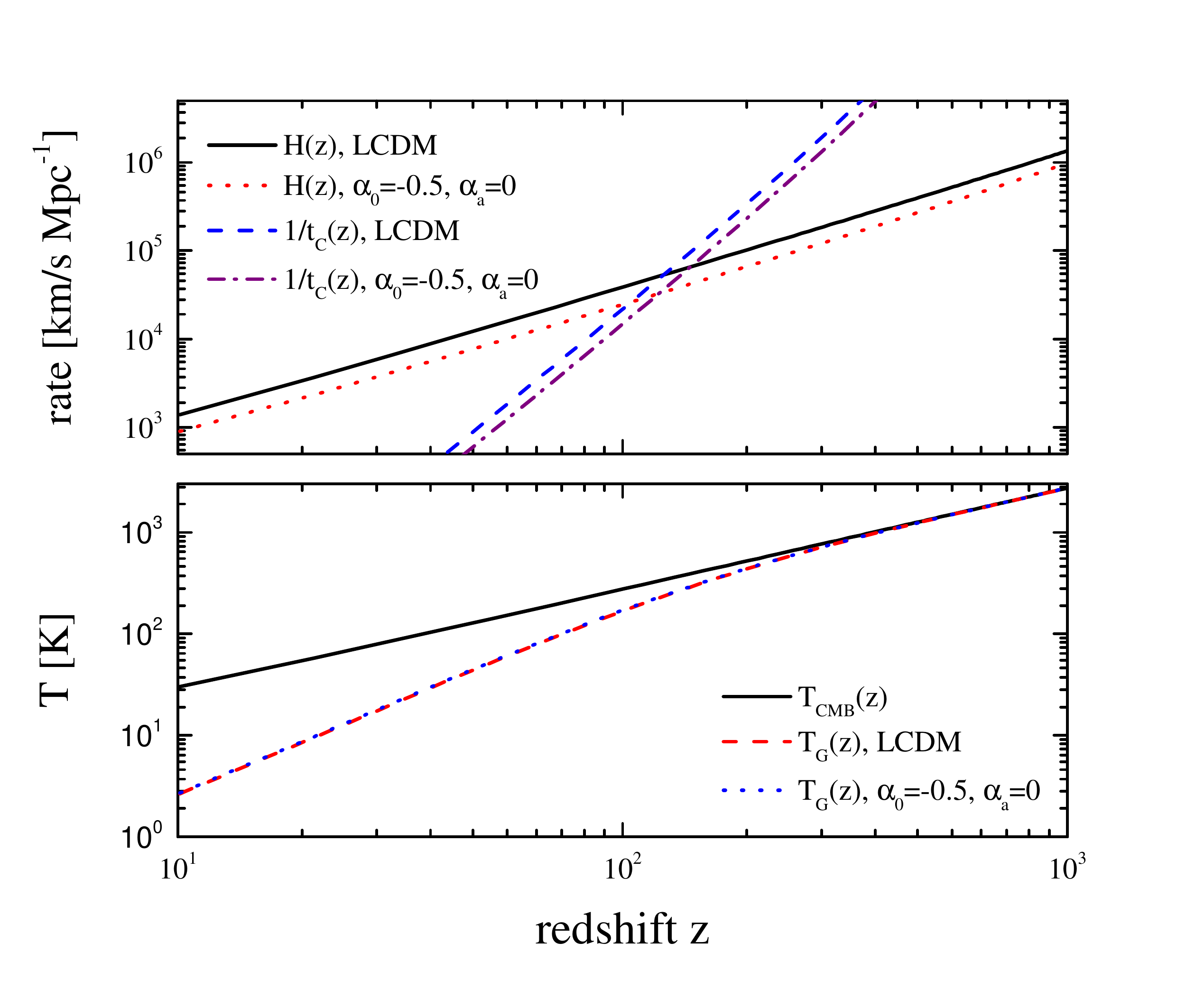}
\caption{Upper panel: The cosmic expansion rate and Compton-heating rate in the $\Lambda$CDM and interacting vacuum energy model with fixed parameters \ie\,$\alpha_0=-0.5$ and $\alpha_a=0$. Lower panel: CMB temperature and gas temperatures in the $\Lambda$CDM and interacting vacuum energy model with fixed parameters \ie\,$\alpha_0=-0.5$ and $\alpha_a=0$.}\label{fig:rate_T_z}
\end{figure} 

\section{Observational constraints} \label{sec:test}
We use a modified version of {\tt CAMB} \citep{CAMB}\footnote{Available at \url{https://camb.info}} to compute the theoretical prediction of $T_{21}(z)$ using Eqs (\ref{eq:T21})-(\ref{eq:H}), given a set of cosmological parameters,
\ba\label{eq:P}
P \equiv \{\omega_b, \omega_c, \Theta_s, \tau, n_s, A_s,\alpha_0, \alpha_a\}\,,
\ea
where $\omega_b$ and $\omega_c$ are the physical baryon and CDM densities respectively, $\Theta_s$ is 100 $\times$ the ratio of the sound horizon to the angular diameter distance at decoupling, $\tau$ is the reionization optical depth, $n_s$ and $A_s$ are the spectral index and the amplitude of the primordial power spectrum respectively, and $\alpha_0$ and $\alpha_a$ parametrise the strength of the interacting vacuum in the form of Eq. (\ref{eq:alpha}).

It is assumed that the spin temperature $T_{\rm S}(z)$ fully couples to the gas temperature $T_{\rm G}(z)$ at redshifts $z\simeq15-20$, as indicated by the observed 21-cm signal from EDGES, and as discussed in recent paper \citep{Xiao:2018jyl}, and we compute the evolution of $T_{\rm G}(z)$ using {\tt RECFAST} \citep{Seager:1999bc, Seager:1999km, Wong:2007ym, Scott:2009sz}\footnote{Available at \url{http://www.astro.ubc.ca/people/scott/recfast.html}}. The evolution equation of the gas temperature $T_{\rm G}(z)$ is given in \citep{Seager:1999bc,Scott:2009sz}, \ie\,
\ba\label{eq:TG}
\frac{dT_{\rm G}(z)}{dz}=\frac{T_{\rm G}(z)-T_{\rm CMB}(z)}{H(z)(1+z)t_C(z)} + \frac{2T_{\rm G}(z)}{1+z}\,,
\ea
where $t_C(z)$ is the Compton-heating timescale, \ie\,
\ba
t_C(z)= \frac{3 m_e c}{8 \sigma_T a_R T^4_{\rm CMB}(z) } \left[ \frac{1+f_{\rm He}(z) +x_e(z)}{x_e(z)} \right] \,.
\ea
Here $m_e$ is the electron mass, $c$ is the speed of light, $\sigma_T$ is the Thomson scattering cross section, $a_R$ is the radiation constant, $f_{\rm He}(z)$ is the fractional abundance of helium by number, and $x_e(z)$ is the free electron fraction normalized to the total hydrogen number density. The decoupling time between gas and CMB is at $H\approx1/t_C(z)$.

As shown in Eq. (\ref{eq:TG}), the evolution of gas temperature also depends on $H(z)$. In order to figure out the effect of $H(z)$ on the EDGES signal, we show the cosmic expansion rate $H(z)$ and Compton-heating rate $1/t_C(z)$ in the $\Lambda$CDM and interacting vacuum energy model with fixed parameters \ie\,$\alpha_0=-0.5$ and $\alpha_a=0$ in the upper panel of Fig. \ref{fig:rate_T_z}. It is seen that the interacting vacuum model has a smaller $H(z)$, but the decoupling time between gas and CMB at which $H\approx1/t_C(z)$ in the interacting vacuum energy model has very little change, compared with that in $\Lambda$CDM. In the lower panel of Fig. \ref{fig:rate_T_z}, we show the CMB temperature and gas temperatures in the $\Lambda$CDM and interacting vacuum energy model with fixed parameters \ie\,$\alpha_0=-0.5$ and $\alpha_a=0$. Both models have very close gas temperatures. Therefore, according to Eq. (\ref{eq:T21}) we can see that a reduced value of $H(z)$ in Eq.  (\ref{eq:tau}) would be the main contribution to an increase of the amplitude of 21-cm signal.

\begin{figure}[!t]   
\includegraphics[scale=0.3]{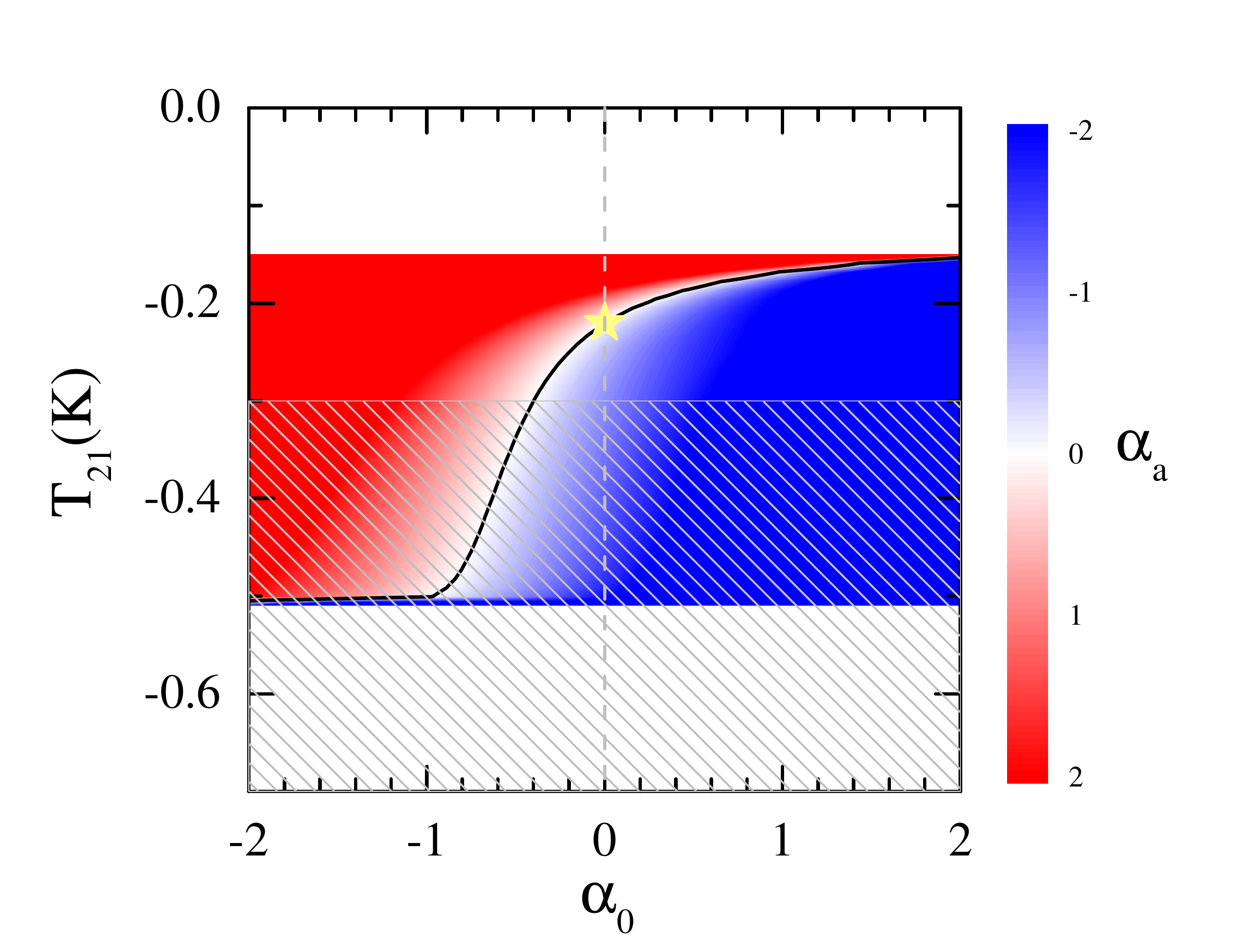}
\caption{An illustration of the intensity of the 21-cm signal relative to the CMB temperature, $T_{21}(\rm K)$, for various values of $\alpha_0$ and $\alpha_a$. The colour bar indicates the values of $\alpha_a$. The black solid curve corresponds to the case in which $\alpha$ does not evolve with time ($\alpha_a=0)$. The intersect between the black solid curve and the vertical grey dashed line, marked by a yellow star, denotes the $\Lambda$CDM model. The hatched region illustrates the observed 21-cm signal from EDGES at 99\% CL.}\label{fig:T21_alpha0}
\end{figure}

In Fig. \ref{fig:T21_alpha0}, we show $T_{21}$, as defined in Eq (\ref{eq:T21}), for various values of $\alpha_0$ and $\alpha_a$, with other cosmological parameters fixed at values consistent with a Planck 2015 cosmology \citep{Ade:2015xua}. As shown, the $\Lambda$CDM model ($\alpha_0=\alpha_a=0$), denoted by the star, is in tension with the EDGES measurement at 99\% CL illustrated by the hatched region. However, interacting vacuum models can in principle release the tension, namely, $T_{21}$ can be pushed into the hatched region by a large range of the $\alpha_0$ and $\alpha_a$ parameters.

\begin{figure}[b!]   
\includegraphics[scale=0.3]{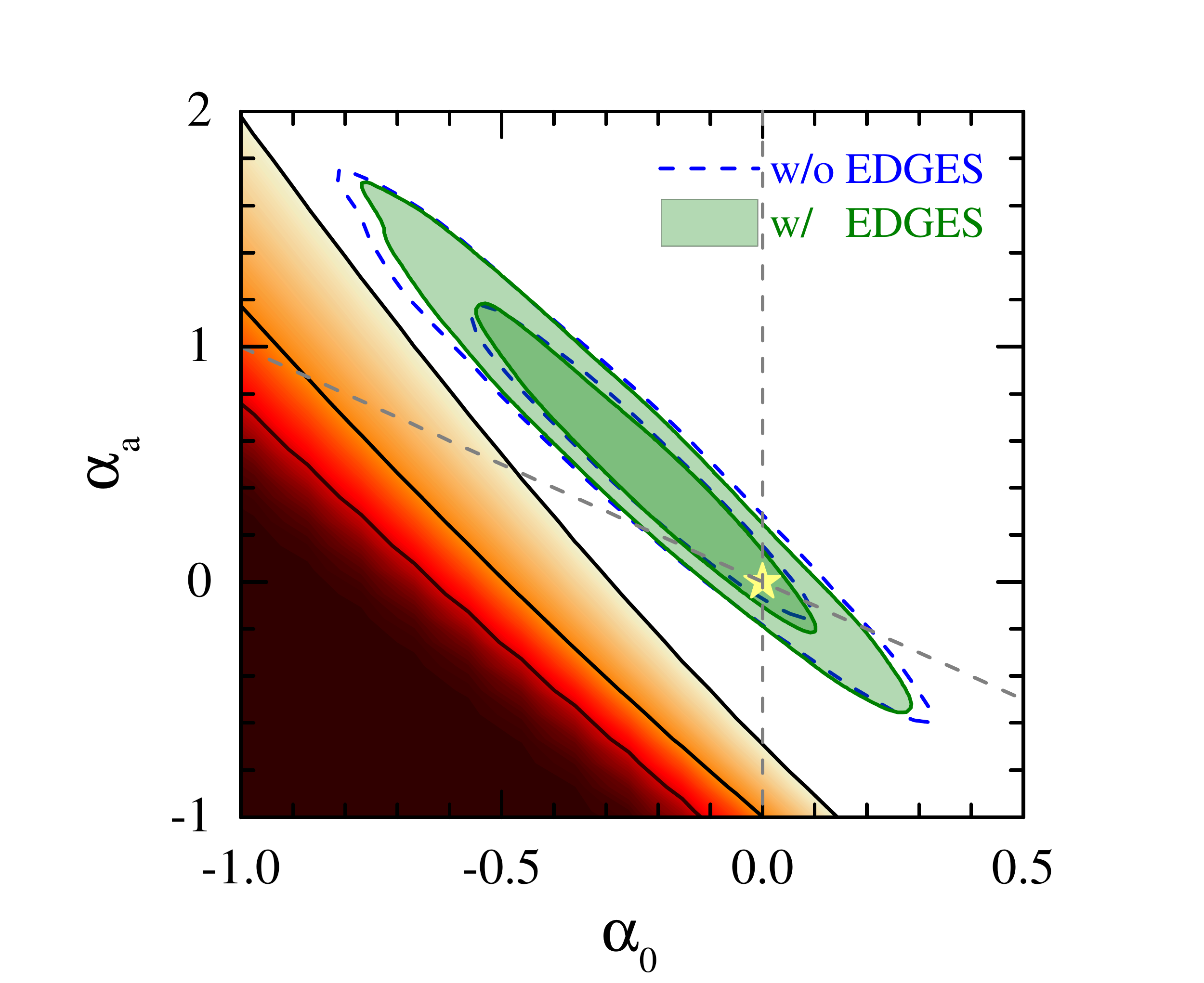}
\caption{The contour plots for parameters $\{\alpha_0, \alpha_a\}$ derived from different data combinations including EDGES alone (shaded regions and solid curves in the left corner; the solid curves from left to right denote 68, 95 and 99\% CL contours respectively), CMB + SNe + BAO + RSD + $H_0$ (blue dashed), and CMB + SNe + BAO + RSD + $H_0$ + EDGES (solid green). The yellow star marks the $\Lambda$CDM model. The dashed lines denote $\alpha_0=0$ and $\alpha_0+\alpha_a=0$, which divide the $\alpha_0$-$\alpha_a$ parameter space into four regions, as illustrated in Fig. \ref{fig:scalar_field_like}.}\label{fig:contour_q1_q2}
\end{figure} 

We then perform a Monte Carlo Markov Chain (MCMC) global fit for parameters in Eq (\ref{eq:P}) using a modified version of {\tt CosmoMC} \footnote{Available at \url{https://cosmologist.info/cosmomc/}} \citep{Lewis:2002ah} with a combined dataset including,

\begin{itemize}
\item  The angular power spectra of temperature and polarization measurements of CMB from the Planck mission \citep{Aghanim:2015xee};
\item  The Joint Light-curve Analysis (JLA) sample of supernovae (SNe) measurements \citep{Betoule:2014frx};
\item  The Baryonic Acoustic Oscillations (BAO) distance measurements from 6dFGS \citep{6df}, SDSS DR7 Main Galaxy Sample \citep{MGS},  Lyman-$\alpha$ forest of BOSS DR11 quasars \citep{Font-Ribera:2013wce, Delubac:2014aqe}, BOSS DR12 with tomographic information \citep{BAOWang, BAOZhao}; joint BAO and Redshift Space Distortions (RSD) measurements from WiggleZ \citep{Blake:2012Apr} and from eBOSS DR14 \citep{Zhao:2018jxv}; and RSD measurements from 6dFGS \citep{Beutler:2012Apr}, 2dFGRS \citep{Percival:2004Jun}, SDSS LRG \citep{Samushia:2011Feb} and VIPERS \citep{Torre:2013Mar};
\item  The local $H_0$ measurement using the Cepheids, \ie\, $H_0=73.24 \pm 1.74 \rm \,km\,s^{-1}\,Mpc^{-1}$  \citep{Riess:2016jrr}.
\end{itemize}

\begin{figure}[tbp]   
\includegraphics[scale=0.25]{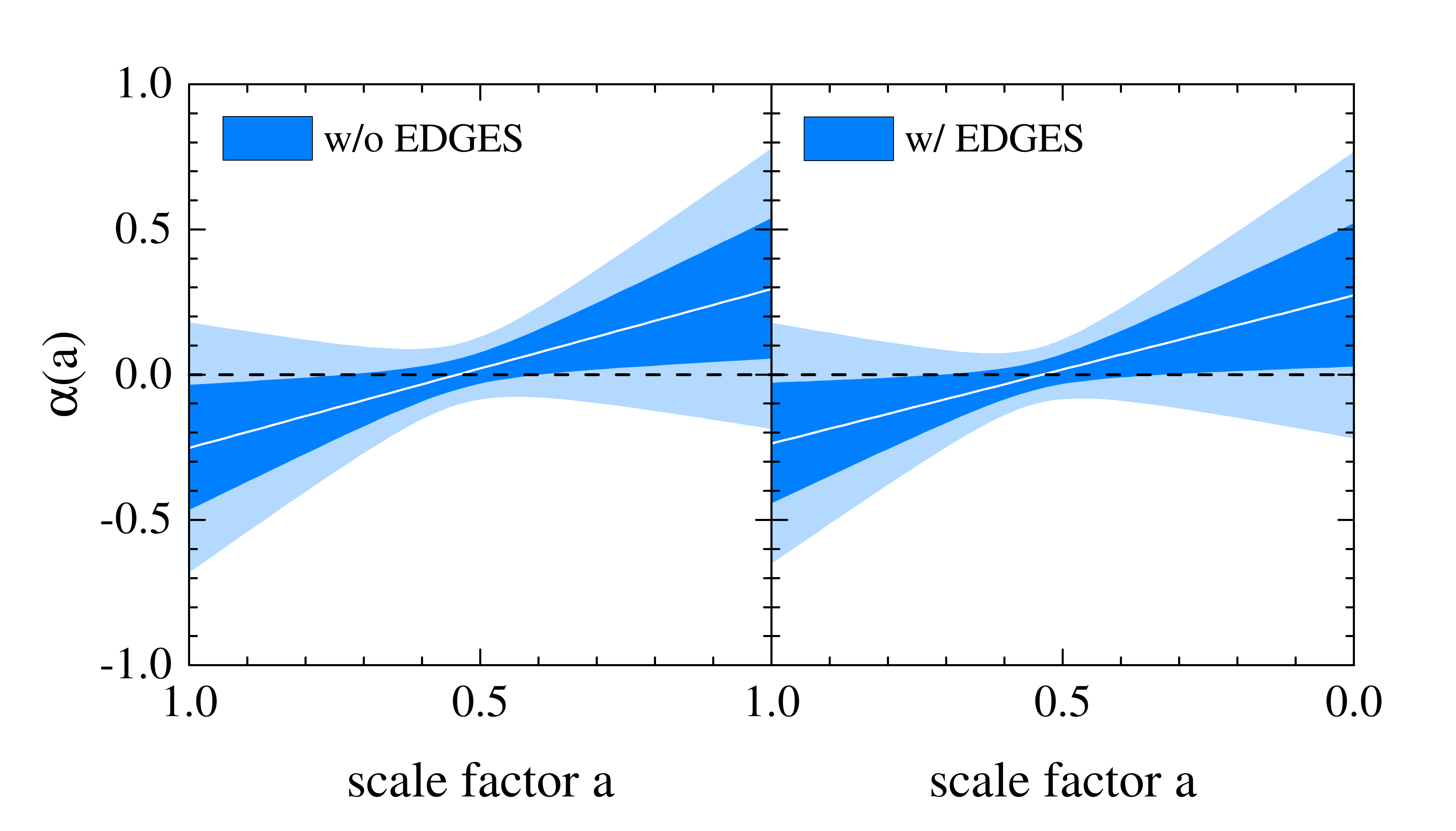}
\caption{The 68 and 95\% CL parametric reconstruction of $\alpha(a)$ using CMB+SNe+BAO+RSD+$H_0$ with (right) and without (left) the EDGES measurement.}\label{fig:alpha_a} 
\end{figure} 

\begin{table}
\begin{center}
\begin{tabular}{c|c|c}
  \hline
      \hline
            & \multicolumn{2}{c}{CMB+SNe+BAO+RSD+$H_0$} \\
   & \multicolumn{1}{c|}{w/o EDGES}&\multicolumn{1}{c}{w/ EDGES}\\
  \hline
  $\alpha_0$ & $-0.252\pm0.216 $ & $-0.237\pm0.208 $\\
  $\alpha_a$ & $ ~~0.547 \pm 0.446 $ & $~~0.510 \pm 0.445 $ \\
  \hline
  FoM            & $1$ & $1.1$ \\
  \hline
    \hline
\end{tabular}
\caption{The mean and 68\% CL constraints on parameters for interacting vacuum (with all other relevant parameters marginalised over) using the combined datasets of CMB, SNe, BAO, RSD and $H_0$ with or without the EDGES measurement. The last row shows the FoM for $\alpha_0$ and $\alpha_a$ (the case without EDGES is normalised to 1).}
\label{tab:result}
\end{center}
\end{table}

The results are summarised in Table \ref{tab:result} and in Figures \ref{fig:contour_q1_q2} and \ref{fig:alpha_a}. To   quantify the constraint on $\alpha_0$ and $\alpha_a$ from the EDEGS measurement, we show the 68, 95 and 99\% CL contours of $\alpha_0$ and $\alpha_a$ using EDEGS alone in the lower left part of Figure \ref{fig:contour_q1_q2} \footnote{As EDEGS alone cannot constrain all parameters in Eq (\ref{eq:P}) simultaneously, we only vary $\alpha_0$ and $\alpha_a$ in this case while other parameters are fixed to the values derived from the Planck 2015 measurement.}. As illustrated, the EDGES measurement favours an interacting vacuum model over the $\Lambda$CDM model at more than 99\% CL. And this would induce a quintessence-like effective dark energy as shown in the lower left of Figure \ref{fig:scalar_field_like}. We then perform a global fit for all parameters in Eq (\ref{eq:P}) using the above-mentioned joint dataset with and without the EDEGS measurement, and show the 68 and 95\% CL contours of $\alpha_0$ and $\alpha_a$ in Figure \ref{fig:contour_q1_q2}. As we can see, the EDGES data make the contours shrunk marginally without changing the degeneracy between $\alpha_0$ and $\alpha_a$, namely, the error bars of $\alpha_0$ and $\alpha_a$ get tightened by 4\% and 0.2\% respectively, and the Figure of Merit (FoM), which is the determinant of the inverse covariance matrix for $\alpha_0$ and $\alpha_a$, gets improved by 10\%, as shown in Table \ref{tab:result}. The $\Lambda$CDM model is compatible with data within 68\% CL in both cases. This is expected as the comparatively low precision of the EDGES measurement makes it difficult to compete with the remaining combined datasets.

To see the evolution history of $\alpha$ allowed by current observations, we reconstruct $\alpha(a)$ using the constraint that we derived with the functional form assumed in the first place, and show the result in Figure \ref{fig:alpha_a}. As expected, adding EDGES data barely changes the reconstruction, and a sweet spot, the epoch at which the error of $\alpha$ gets minimised, shows up at $a\sim0.5$ ($z\sim1$) in both cases, and at this epoch, the best-fit value of $\alpha$ changes sign \ie, energy transfers from dark matter to vacuum energy at early times ($z\gtrsim1$), and {\it vice versa} at late times ($z\lesssim1$).

\section{Conclusion}

The recent measurement of the 21-cm brightness temperature performed by the EDGES team has attracted wide attention, partially due to the fact that the observed signal is far below what is expected in a $\Lambda$CDM model. Interpretations have been proposed, and most of which focus on the nature of dark matter. 

In this work, we perform a proof-of-the-concept study of the potential power of 21-cm measurements to constrain the possible interaction between dark matter and dark energy. We find that EDGES alone can yield a non-trivial constraint on $\alpha_0$ and $\alpha_a$, parameters quantifying the interaction (with all other parameters fixed), and an interaction vacuum model is able to explain the measured 21-cm brightness temperature. 

Given the large uncertainty in the current EDGES measurement, it marginally improves the constraint on $\alpha_0$ and $\alpha_a$ on top of a compilation of recent measurements of SNe, CMB, BAO and RSD, and the $\Lambda$CDM model agrees with the combined datasets within 68\% CL. An improved test will be benefited from more realistic systematic uncertainties in the furture. Additionally, future 21-cm measurements, such as the square kilometre array (SKA)\footnote{More information is available at \url{https://www.skatelescope.org/}}, will provide much more precise measurement on $\alpha$, which offers a new probe to shed light on nature of dark energy and dark matter.

\begin{acknowledgments}
We thank Bin Yue and Xuelei Chen for discussions. We also thank David Wands for discussions and comments. YW and GBZ is supported by NSFC Grants 1171001024 and 11673025. GBZ is also supported by the National Key Basic Research and Development Program of China (No. 2018YFA0404503) and a Royal Society Newton Advanced Fellowship, hosted by University of Portsmouth. YW is also supported by the Nebula Talents Program of NAOC and the Young Researcher Grant of NAOC. This research used resources of the SCIAMA cluster supported by University of Portsmouth, and the ZEN cluster supported by NAOC.

\end{acknowledgments}



\end{document}